# scientific data

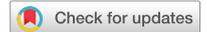

**OPEN**

**DATA DESCRIPTOR**

# A Real-World Energy Management Dataset from a Smart Company Building for Optimization and Machine Learning

Jens Engel[1,2 ✉], Andrea Castellani[1], Patricia Wollstadt[1], Felix Lanfermann[1], Thomas Schmitt[1], Sebastian Schmitt[1], Lydia Fischer[1], Steffen Limmer[1], David Luttropp[1], Florian Jomrich[3], René Unger[4] & Tobias Rodemann[1]

We present a large real-world dataset obtained from monitoring a smart company facility over the course of six years, from 2018 to 2023. The dataset includes energy consumption data from various facility areas and components, energy production data from a photovoltaic system and a combined heat and power plant, operational data from heating and cooling systems, and weather data from an on-site weather station. The measurement sensors installed throughout the facility are organized in a hierarchical metering structure with multiple sub-metering levels, which is reflected in the dataset. The dataset contains measurement data from 72 energy meters, 9 heat meters and a weather station. Both raw and processed data at different processing levels, including labeled issues, is available. In this paper, we describe the data acquisition and post-processing employed to create the dataset. The dataset enables the application of a wide range of methods in the domain of energy management, including optimization, modeling, and machine learning to optimize building operations and reduce costs and carbon emissions.

## Background & Summary

Buildings make up a large share of the global energy consumption, representing approx. 40% of primary energy consumption in the EU and US[1]. As the transition towards renewable energy production is ongoing, this sector will thus have a large leverage on shaping this transition. This transition is driven by the adoption of new technologies, such as the installation of new sensors and control components, the use of energy management systems (EMSs) or the deployment of energy forecasting. The realization of all these aspects requires informed decision making[2,3]. Therefore, detailed data on energy consumption of a large variety of buildings is needed[4–7]. However, such data is only sparsely available[5,7]. One reason for this is that collecting large amounts of data is a complex and expensive task, as it requires setting up substantial physical (installation of meters) as well as IT infrastructure (data collection and processing). Another reason is that even when data is being collected, it is often not made publicly available. Especially rare is high quality data, as data collection is an error-prone process, and correction of issues requires active monitoring and problem resolution.

In this paper, we present a comprehensive, curated dataset containing six years of energy usage data of a medium-sized industrial facility located in Offenbach am Main, Germany. The data was recorded from January 1, 2018 to December 31, 2023, thus including normal operation as well as reduced office occupation during the COVID-19 pandemic. The facility consists of offices, workshops, server rooms, and a vehicle emissions lab. It further harbors a 749 kW$_p$ photovoltaic (PV) system as well as a gas-fired combined heat and power plant (CHP) with 199 kW$_{el}$. The building's heating, ventilation, and air conditioning (HVAC) system is supplied with heat

[1]Honda Research Institute Europe GmbH, Carl-Legien-Str. 30, 63073, Offenbach am Main, Germany. [2]Control Methods and Intelligent Systems Laboratory, Technical University of Darmstadt, Landgraf-Georg-Strasse 4, 64283, Darmstadt, Germany. [3]Honda R&D Europe (Deutschland) GmbH, Carl-Legien-Str. 30, 63073, Offenbach am Main, Germany. [4]EA Systems Dresden GmbH, Würzburger Str. 14, 01187, Dresden, Germany. ✉e-mail: jens.engel@honda-ri.de

from the CHP and gas-fired heating boilers, while cooling power is mainly provided by a central cooling system consisting of three water chillers. The dataset contains hierarchical and structured measurement data of

- all electrical components,
- the central heating and cooling systems,
- a roof-mounted weather station.

Unique features of the presented dataset include

- 6 years of structured, hierarchical meter data of electrical and thermal systems,
- labeling of manually specified and automatically detected issues found in the meter data,
- a thorough correction and labeling of all detected issues,
- raw and corrected data at different processing steps,
- detailed electrical meter data containing energy, power, voltage, current, power factor, and frequency measurements at 1 min, 15 min and 1 h resolution,
- an aggregated version of the main dataset for easy evaluation, containing the main electrical and thermal consumption meters and weather data.

The presented dataset is relevant for a range of applications and has been used prior to publication in various use cases, namely

- development and evaluation of EMSs[8–11],
- energy system optimization[12–16],
- data-driven modeling and control[17–19],
- anomaly detection[20,21],
- load prediction and disaggregation[22].

In addition, the data has been used for monitoring purposes at the facility in the past years. Further applications, e.g., network inference[23–25], are conceivable.

## Methods

In this section, we first describe the building with its facilities and components. Then we describe how measurement data was collected, cleaned, and processed.

**Description of the building and components.** The facility at which the data has been collected is the Honda R&D Europe facility, located in Offenbach am Main, Germany. Construction of the building began in 1990 and was completed in 1992. The facility is connected to the power grid through four 20 kV to 230 V two-winding transformers, which are operated pairwise in parallel, as illustrated in Fig. 1. The HVAC is supplied with heat by two natural gas boilers with a total heating power of approximately 1500 kW and a gas-fired CHP. The CHP is a natural gas powered Viessmann Vitobloc 200 with 199 $kW_{el}$ and a power-to-heat ratio of 0.677. The CHP is used if the heat demand is above a certain threshold and the return temperatures are low enough for safe operation.

Both the CHP's electrical and thermal production are metered. The central heating power is measured using two meters: a central meter, measuring the total heating power production from both the boiler and the CHP, and one meter for the heat production of the CHP, as illustrated in Fig. 2. Central cooling is provided by three water chiller cooling machines, CM1, CM2, and CM3, with a combined total cooling power of approx. 1550 kW. The electrical consumption of these cooling machines is measured with one (CM1) and two (CM2 and CM3) electricity meters per cooling machine, respectively, while the thermal cooling power is metered with various meters throughout the building. The electrical power of the recooler fans is measured separately as part of the overall ventilation system. The schematic of the central heating and cooling meters is shown in Fig. 2. The cooling power is metered using various submeters, including a high-level meter, which measures the total cooling power production, and submeters in the subdivisons of the HVAC system.

Apart from to the central cooling system illustrated in Fig. 2, additional local cooling machines are installed in the design studio of the building, as well as for the servers in the workshop areas. However, only their electrical consumption is metered, as shown in Fig. 1 (meters `H3.Z45`, `H2.Z(E)66` and `H2.Z(E)67`). Furthermore, local split air conditioners are installed throughout parts of the facility, which have neither thermal nor individual electrical metering. Electrical consumption of the ventilation system is metered using seven separate meters as illustrated in Fig. 1. For more details on the thermal aspects of the building and zones, the reader is referred to Schmitt, T. 2022[26], which provides a simplified linear thermal and electrical model description of the facility.

The building further has a PV system, consisting of six groups of modules, totaling in a peak production capability of 749 $kW_p$. The parameters of these six groups are shown in Table 1 and their placement on the facility is illustrated in Fig. 3. The PV system was installed in two phases: Groups 1 and 2 were installed in June 2019, while groups 4–6 followed in June 2020. The PV system is metered using four meters according to the physical distribution of the panels across the roofs, and hence not by orientation of the panels, as listed illustrated in Fig. 1. Groups 1 and 2 are metered by `V.Z84/V.ZE84`. Groups 3 and 4 are metered by three separate meters: 18.8% of the panels, which are located on the far west of the facility's roof are metered by `H1.Z310`, the 57.8% mounted on the central roofs by `H2.Z311`. The remaining 23.4% of the panels on the eastern part of the building are metered together with groups 5 and 6 by `H3.Z312`.

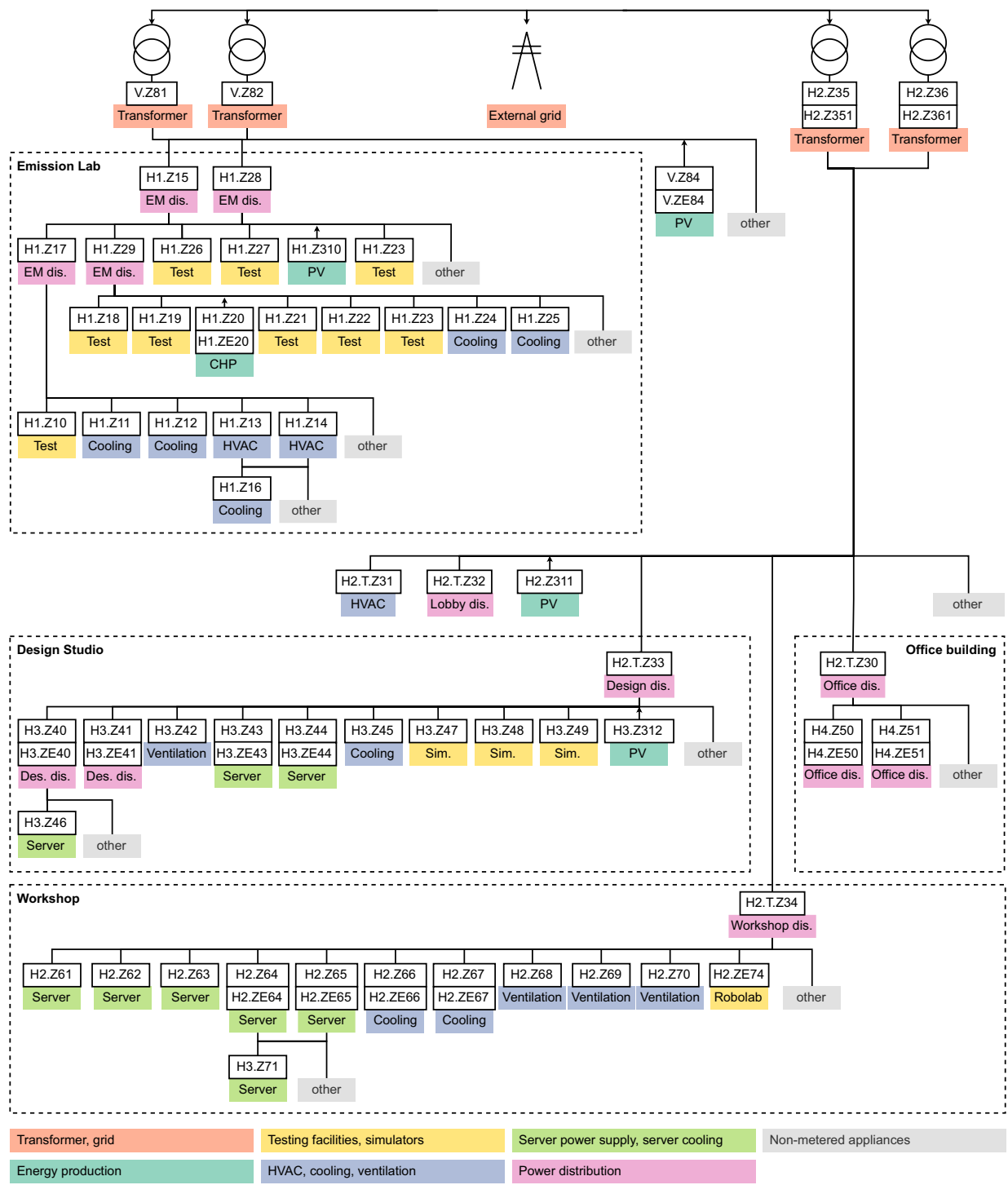

**Fig. 1** Schematic of the electricity metering hierarchy of the facility. Each sensor is labeled with a uniform resource name (URN) which reveals some information on the sensor: H1, H2, H3, H4 and V corresponds to the physical location of the meter within the facility, i.e., the sub-distribution it belongs to. As it is to be expected for a real facility where the setup was installed incrementally over a long period of time, this naming scheme is not fully consistent and unified.

For measuring weather information, a pole-mounted weather station is installed on the highest point of the facility, as marked in Fig. 3. The weather station is located at 50.08478° N, 8.84013° E.

**Data collection.** The dataset contains measurements from electricity meters, heating and cooling meters, and the weather station. Electricity meters installed in the facility are ABB-B24, Janitza UMG 96 RM-E, Janitza UMG 96 PA MID+, as well as Socomec DIRIS I35, I45 and S135 meters. Heating and cooling is metered using

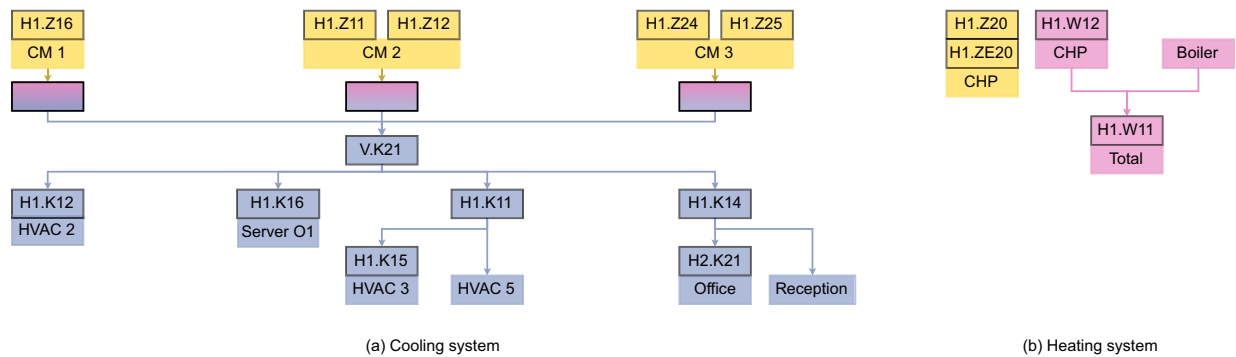

**Fig. 2** Schematic of the central cooling- and heating-system with respective meters. Yellow nodes indicate electricity meters, blue nodes cooling meters, pink nodes heating meters. The nodes immediately below the cooling machines indicate the heat exchangers of the cooling machines. The CHP's co-produced electricity is redundantly metered by two meters.

| Group | Module Type | Facing | Slope | Peak Power |
|---|---|---|---|---|
| 1 | SI-Enduro 300 $W_p$ | 95° East | 5° | 68 $kW_p$ |
| 2 | SI-Enduro 300 $W_p$ | 275° West | 5° | 68 $kW_p$ |
| 3 | Heckert Nemo 2.0 325 $W_p$ | 70° East | 11° | 274 $kW_p$ |
| 4 | SI-Enduro 300 $W_p$ | 250° West | 11° | 274 $kW_p$ |
| 5 | Heckert Nemo 2.0 325 $W_p$ | 58° East | 11° | 32.5 $kW_p$ |
| 6 | Heckert Nemo 2.0 325 $W_p$ | 238° West | 11° | 32.5 $kW_p$ |

**Table 1.** Parameters of the 6 individual groups making up the PV system of the facility.

SensorStar 2/2U meters. The list of all meters and uniform resource names (URNs) present in the dataset can be found in Tables 2 and 3. Weather measurements are collected from a Lufft WS501-UMB weather station.

Data from the meters and the weather station is queried and processed by a number of different IP gateways, namely Tixi Data Gateways, Socomec DIRIS D50 and D70, and VisualGateway gateways, a proprietary gateway solution for querying Janitza meters and the weather station. The gateways query the various meters via Modbus/RTU or M-Bus protocol. From the gateways, data is transfered via HTTPS to a proprietary monitoring server, which in turn stores the data into an influxdb time series database. Data from the Janitza and Socomec meters, as well as the weather station, is recorded only by change-of-value, i.e., only when the measured values change by more than a certain threshold. For ABB-B24 and SensorStar meters however data is recorded periodically at 1 min sample resolution. Both Tixi Data Gateways and VisualGateway gateways buffer queried data locally. The buffer is queried by the monitoring server at regular intervals to transfer the data into the influxdb. The buffer can hold data of an extended amount of time, and can thus bridge temporary communication failures. The Socomec gateways have no such buffer and are queried directly from the monitoring server. While the Tixi gateways query data continuously, they transfer data into the buffer only once every 1 min, at which all collected data within this interval is transmitted. This behavior introduces a jitter in the recorded data of 0–59 s which cannot be mitigated. As the gateways query and collect measurement data, they attach UTC timestamps based on their internal device time. Initially, until March 2020, only the VisualGateway gateways synchronized their internal time through a time server. The Tixi gateways do not support time synchronization via the network time protocol (NTP) and were thus not synchronized, leading to drift in the timestamps, rendering correlation or summation of measurements across different gateways difficult. Since March 2020, the internal time of the Tixi gateways has been synchronized periodically by updating it via the gateways' diagnostics interface through the monitoring server. The drift present in the data prior to this date is systematic and was mitigated in the dataset. Since the Socomec meters are queried directly by the monitoring server, their measurements received synchronized timestamps.

The heating and cooling meters collected measurements of power, energy, volume flow, and flow temperatures. The employed meters calculate power and energy from temperature and volume flow measurements. The measurements of the heating and cooling meters are listed in Table 4. For all electricity meters, power, energy, frequency information as well as three-phase current, voltage, power and power factor were collected. All measurements of the electricity meters are listed in Table 5. The measurements recorded by the weather station are listed in Table 6.

**Data cleaning and post-processing.** The presented dataset was recorded over the course of 6 years between January 1, 2018 until December 31, 2023. During this time span, a multitude of issues occurred which directly affected the data collection, like measurement outages, maintenance, device replacements and addition of new sensors.

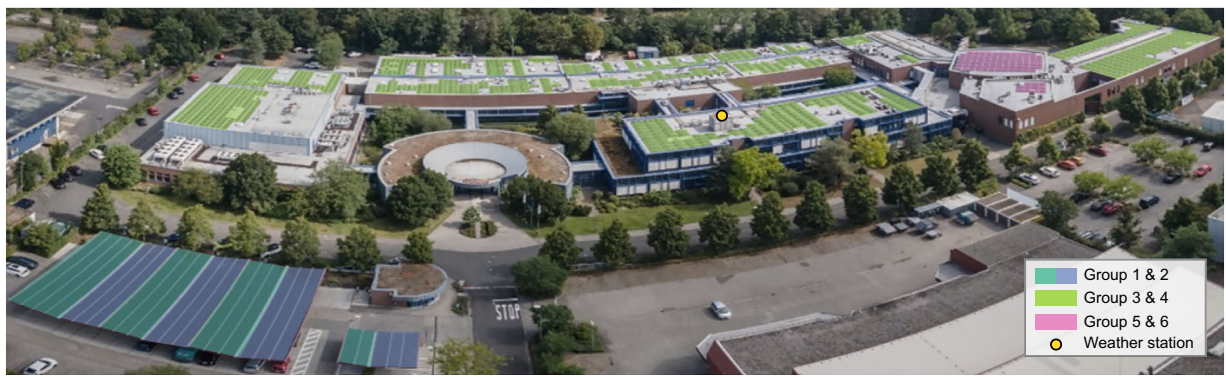

**Fig. 3** Aerial view of the Honda R&D facility in Offenbach, Germany with highlighted PV groups. PV modules are grouped by orientation, and two groups of opposite orientation each are installed together. The pole-mounted weather station is located in the center of the facility to deliver representative measurements of solar irradiation.

In order to produce a consistent dataset, the impact of such issues needs to be clarified and the corresponding data corrected, whenever possible. We therefore apply a cleaning and post-processing pipeline to the data, which is illustrated in Fig. 4. In a first step, known issues are manually specified. Additionally, systematic issues are detected using a rule-based detection mechanism. Before correcting the manually and automatically collected issues, first, data is harmonized to ensure consistency in naming and sign convention. After issue correction is applied, all measurements are aligned in time before being resampled into equidistantly sampled time series. Eventually, based on the resampled data, single missing measurements are calculated, and processed data is exported. In the following, each step of the pipeline will be described in more detail.

*Manual specification of known issues.* A number of issues that were caused by external influences or installation problems are known. These were specified manually and then corrected automatically later in the pipeline, as will be described in the issue correction section. The following such issues were encountered:

**Main cooling meter failure.** The mechanical flow sensor of the main cooling meter `V.K21`, which meters the cooling power generated by all three cooling machines, broke twice for extended periods of time. Measurements are reconstructed from submeters, as described in the issue correction section.

**Design studio meters powered off.** For an extended period of time (May 14 - June 4, 2018), main and submeters in the design studio were powered off, causing a gap in the data.

**Wrongly configured design studio main meter.** The main meter of the design studio `H2.T.Z33` was configured wrongly during January 17 – 29, 2018, yielding faulty measurement data. This time span was marked for removal from the dataset, yielding a gap to be corrected, as described in the issue correction section.

**Implausible PV meter readings.** PV meter `H2.Z311` recorded implausible power/energy measurements during November 12 – 16, 2020, hence the time span was marked for removal.

**Wrong conversion factor in meter.** After installation of the PV system in June 2020, the meter `H2.Z311`, which is connected through a current transformer, was configured with an incorrect conversion factor accounting for the transformation, until correction in February 2021. During this time span, the measurements of power, energy and current need to be rescaled. Similarly, after installation of the redundancy meters `H4.ZE50` and `H4.ZE51`, incorrect conversion factors were configured, which has not yet been corrected in the physical installation. Therefore, all corresponding measurements need to be rescaled.

**Interference on test chamber meter `H1.Z19` from nearby CHP wire.** Since maintenance in November 2022, the line measured by meter `H1.Z19` was unused, yet measured non-zero power due to interference from the nearby CHP line. Between March and November 2022, after installation of nearby CHP meter `H1.ZE20`, the meter `H1.Z19` incorrectly measured asymmetric loads across the three phases. Therefore, the period from March 2022 until December 2023 was marked for removal.

*Rule-based issue detection.* Most measurement issues can be classified as one of three types: Zero value measurements, value leaps or measurement gaps. These issues can be detected reliably using simple rules:

**Zero measurement detection.** A zero value in a cumulative measurement (i.e., energy or flow volume) right after a non-zero value indicates a measurement issue. The issue is marked from the beginning of the zero value until it changes back to a non-zero value. This issue occurs when gateways or meters are restarted and measurement variables are initialized with zero or null until regular operation resumes. In the presented dataset, this occurs mostly with meters monitored by Tixi Data Gateways.

| URN | Type | Description |
|---|---|---|
| **Electricity Main** | | |
| V.Z82 | Janitza UMG 96 PA MID+ | Parking lot transformer 1 |
| V.Z81 | Janitza UMG 96 PA MID+ | Parking lot transformer 2 |
| H2.Z35 | ABB-B24 | Office transformer 1 |
| H2.Z351 | Janitza UMG 96 PA MID+ | Office transformer 1 replacement |
| H2.Z36 | ABB-B24 | Office transformer 2 |
| H2.Z361 | Janitza UMG 96 PA MID+ | Office transformer 2 replacement |
| **Local Generators** | | |
| H1.Z(E)20 | ABB-B24/Janitza UMG 96 PA MID+ | CHP production |
| V.Z(E)84 | Socomec I35/Janitza UMG 96 PA MID+ | PV Parking Lot |
| H1.Z310 | Janitza UMG 96 PA MID+ | PV Emission Lab |
| H2.Z311 | Janitza UMG 96 PA MID+ | PV Office Building |
| H3.Z312 | Janitza UMG 96 PA MID+ | PV Design Studio |
| **Servers** | | |
| H3.Z(E)43 | ABB-B24/Janitza UMG 96 PA MID+ | Server O4 cooling 1 |
| H3.Z(E)44 | ABB-B24/Janitza UMG 96 PA MID+ | Server O4 cooling 2 |
| H3.Z46 | ABB-B24 | Server O4 power supply |
| H2.Z61 | ABB-B24 | Server CIS local cooling |
| H2.Z62 | ABB-B24 | Server CIS power supply 1 |
| H2.Z63 | ABB-B24 | Server CIS power supply 2 |
| H2.Z(E)64 | ABB-B24/Janitza UMG 96 PA MID+ | Server EU power supply 1 |
| H2.Z(E)65 | ABB-B24/Janitza UMG 96 PA MID+ | Server EU power supply 2 |
| H2.Z(E)66 | ABB-B24/Janitza UMG 96 PA MID+ | Local cooling (1) |
| H2.Z(E)67 | ABB-B24/Janitza UMG 96 PA MID+ | Local cooling (2) |
| H3.Z71 | ABB-B24 | Server O4 power supply |
| **Cooling** | | |
| V.K21 | SensorStar 2C | Main cooling machines 1, 2, 3 |
| H1.K11 | SensorStar 2C | Emission lab HVAC 3/5 |
| H1.K15 | SensorStar 2C | Emission lab HVAC 3 |
| H1.K12 | SensorStar 2C | Emission lab HVAC 1/2 |
| H1.K14 | SensorStar 2C | Emission lab cooling to office |
| H2.K21 | SensorStar 2C | HVAC office |
| H1.K16 | SensorStar 2C | Server room O1 |
| **Heating** | | |
| H1.W11 | SensorStar 2C | Total heat generation |
| H1.W12 | SensorStar 2C | CHP heat generation |
| **Ventilation** | | |
| H2.T.Z31 | ABB-B24 | HVAC office 50/51 |
| H3.Z42 | ABB-B24 | Ventilation design studio |
| H2.Z68 | ABB-B24 | Ventilation System 17 |
| H2.Z69 | ABB-B24 | Ventilation System 16 |
| H2.Z70 | ABB-B24 | Ventilation System 18 |
| **Workshops** | | |
| H2.T.Z34 | ABB-B24 | Feed workshops |
| H2.ZE74 | Janitza UMG 96 PA MID+ | Robolab |

**Table 2.** List of URNs of all meters present in the dataset. For URNs listed with "(E)", two meters with distinct URNs are installed, measuring the same lines. The ZE meters were installed in 2023 to conform with calibration legislation in Germany.

**Leap detection.** In heating and cooling meters, the energy and flow volume measurements can contain leaps. A leap is detected if an implausible value (i.e., exceeding a configured threshold) is measured right after a valid measurement, or if the rate of change between two measurements exceeds a threshold. Leaps can occur due to value overflows within the meters. The time range until it returns to the plausible value range is marked as an issue. If the first measurement after an issue returns to a value close to the last correct measurement, this indicates a simple leap. Otherwise, a so-called lasting leap occurred, meaning the leap incurred a lasting offset that needs to be corrected.

**Gap detection.** A gap is a period without measurements. Since most of the meters collect data only at change-of-value, detecting measurement gaps is not trivial. Therefore, we detect them using a simple heuristic: For all meters, a maximum expected time between measurements can be determined, which can serve as a threshold value. If no measurement was collected for more than the set threshold, a gap issue is marked until a new value is recorded. One specific case of regular gaps is present in the measurement of meters recorded through VisualGateway gateways. These gateways reset periodically every 4 h to avoid memory overflows, resulting in short gaps of 1–4 min at every cycle. This was eventually resolved in June 2020. There are generally two types of gaps: gaps where values were not being transmitted due to sensor outages, gateway failures or network issues, and gaps where the meter got "stuck", i.e., stopped recording incoming data. For electric and thermal energy meters, these can be differentiated: In the first case, data was being collected by the meters but not transmitted, meaning that the reading of the energy measurements before and after the gap are misaligned. In the second case, the meter was offline and did not record any data, hence the value before and after the gap are identical. Due to jitter and some effects during gateway failure, there can also be intermediate cases, where values before and after the gap do not align, but where the difference is not representative. However, the gap detection mechanism does not differentiate between gap types, yet the differentiation is relevant for issue correction, as addressed in the issue correction section. Gaps are detected by applying the described mechanism to a subset of measurements which are expected to continually change. For electricity meters these are $U_1$, $U_2$, $U_3$, $f$, for heating and cooling meters $T_{\text{diff}}$, $T_{\text{vl}}$, $T_{\text{rl}}$, $P$ and for the weather station $Dc$, $S_c$, $T_a$. When a gap was detected in those measurements, it was assumed that the gap applies to all measurements of the respective meter.

*Harmonization of units and signs.* In this step, basic consistency of measurement data is ensured by

- uniform sign convention across all meters (positive values are inflows/consumption, negative values are outflows/production),
- consistent naming of measurements, e.g., renaming $W$ and $W_{\text{exp}}$ to $W_{\text{in/out}}$ for ABB-B24 meters,
- uniform unit conversion to the units listed in Tables 4 and 5.

*Issue correction.* After manual specification and rule-based detection of issues as described above, the issues are automatically corrected using a set of correction mechanisms. The correction pipeline takes the raw time series data and executes the following steps consecutively in the order they are described below. The output time series are contained in the dataset as the `corrected` time series.

**Deletion.** All measurements during time spans marked for deletion are removed, including zero measurements and simple leaps. Removed time spans automatically yield gaps.

**Offset correction.** In the case of a lasting leap, the incurred offset, i.e., the difference between the first valid value after the leap and the last valid value before the leap, is added to all values following the leap.

**Clone gap filling.** This correction mechanism accounts for gaps detected in all measurements, except power and energy measurements of both electricity and heat meters. If a gap's duration

$$d_{\text{gap}} = t_{\text{end}} - t_{\text{start}}, \tag{1}$$

where $t_{\text{start}}$ is the start and $t_{\text{end}}$ the end time of the respective gap, is less than 15 min, it is filled by copying the measurements from $t_{\text{start}} - d_{\text{gap}}$ until $t_{\text{start}}$, which directly precede the gap. Otherwise, we copy data from a corresponding consistent period before the gap. For this, we scroll back in time in weekly steps: We calculate the number of weeks to scroll back,

$$n_{\text{weeks}} = 1 + \left\lfloor \frac{d_{\text{gap}}}{7\text{d}} \right\rfloor, \tag{2}$$

copying the the measurement data from $t_{\text{start}} - n_{\text{weeks}} \cdot 7\text{d}$ until $t_{\text{end}} - n_{\text{weeks}} \cdot 7\text{d}$. We choose this method instead of interpolation, as most measurements affected by long gaps exhibit weekly-regular behavior. For integrative measurements, such as flow volume and $W_{\text{in/out}}$, additional scaling and offsetting is applied to ensure that measurements after the gap are aligned with the filled data. However, the described mechanism is not applied recursively. If the replacement time span itself has a gap, the original gap will not be filled but will remain in the time series. Note that due to the nature of the data, gaps in weather station measurements are not filled.

**Combined power and energy series correction.** For the correction of gaps in energy measurements, it is important to consider that power and energy measurements are connected through

| URN | Type | Description |
|---|---|---|
| **Emission Lab** | | |
| H1.Z15 | Janitza UMG 96 RM-E | Feed emission lab 1 |
| H1.Z28 | Janitza UMG 96 RM-E | Feed emission lab 2 |
| H1.Z17 | Janitza UMG 96 RM-E | Distribution emission lab 1 |
| H1.Z29 | Janitza UMG 96 RM-E | Distribution emission lab 2 |
| H1.Z10 | ABB-B24 | Test chamber 1 |
| H1.Z13 | ABB-B24 | Cooling 1 / HVAC 1.1 |
| H1.Z14 | ABB-B24 | Cooling 1 / HVAC 1.2 |
| H1.Z16 | ABB-B24 | Cooling machine 1 |
| H1.Z11 | ABB-B24 | Cooling machine 2.1 |
| H1.Z12 | ABB-B24 | Cooling machine 2.1 |
| H1.Z24 | ABB-B24 | Cooling machine 3.1 |
| H1.Z25 | ABB-B24 | Cooling machine 3.2 |
| H1.Z19 | ABB-B24 | HVAC test bench 2 |
| H1.Z23 | ABB-B24 | HVAC test chambers |
| H1.Z18 | ABB-B24 | Test chamber 2 CVS |
| H1.Z21 | ABB-B24 | Test chamber 2.1 |
| H1.Z22 | ABB-B24 | Test chamber 2.2 |
| H1.Z26 | Janitza UMG 96 RM-E | Roller test bench |
| H1.Z27 | Janitza UMG 96 RM-E | Roller test chamber |
| **Offices & Distribution** | | |
| H2.T.Z30 | ABB-B24 | Office B2 total |
| H2.T.Z32 | ABB-B24 | Lobby |
| H4.Z(E)50 | ABB-B24/Janitza UMG 96 PA MID+ | Office B4 distribution 3 |
| H4.Z(E)51 | ABB-B24/Janitza UMG 96 PA MID+ | Office B4 distribution 4 |
| **Design Studio** | | |
| H2.T.Z33 | ABB-B24 | Feed design studio |
| H3.Z(E)40 | ABB-B24/Janitza UMG 96 PA MID+ | Design studio distribution 1 |
| H3.Z(E)41 | ABB-B24/Janitza UMG 96 PA MID+ | Design studio distribution 4 |
| H3.Z45 | ABB-B24 | Cooling design studio |
| H3.Z47 | ABB-B24 | Driving simulator general |
| H3.Z48 | ABB-B24 | Driving simulator control |
| H3.Z49 | ABB-B24 | Driving simulator HVAC |
| **Weather station** | | |
| WeatherStation.Weather | Lufft WS501-UMB | Weather station |

**Table 3.** Continued list of URNs of all meters present in the dataset. For URNs listed with "(E)", two meters with distinct URNs are installed, measuring the same lines. The ZE meters were installed in 2023 to conform with calibration legislation in Germany.

$$W(t) = \int_{t_0}^{t} P(\tau) \, d\tau. \tag{3}$$

Therefore, to fill these gaps, first, the replacement power measurements $P(t)$ for all gaps are retrieved through the selection mechanism described in the previous step. Then, for each gap, the integral over the replacement measurements

$$\Delta W_{\text{clone}} = \int_{t_{\text{start}}}^{t_{\text{end}}} P(\tau) \, d\tau, \tag{4}$$

where $t_{\text{start}}$ is the start and $t_{\text{end}}$ the end time of the respective gap, is compared to the difference in energy measured by the meter, i.e.,

$$\Delta W_{\text{meter}} = W_{\text{meter}}(t_{\text{end}}) - W_{\text{meter}}(t_{\text{start}}). \tag{5}$$

If both are within 90% of each other, i.e.,

$$|\Delta W_{\text{meter}} - \Delta W_{\text{clone}}| < 0.9 \cdot |\Delta W_{\text{meter}}| \tag{6}$$

or

| Measurement | Unit | Description |
|---|---|---|
| $P$ | W | Heating/cooling power |
| $W$ | kWh | Total energy |
| $T_{vl}$ | °C | Flow temperature |
| $T_{rl}$ | °C | Return temperature |
| $T_{diff}$ | mK | Temperature difference between flow and return |
| $q_v$ | L/h | Volume flow |
| $V$ | L | Cumulated volume |

**Table 4.** Measurements collected from all thermal (heating/cooling) meters.

| Measurement | Unit | Description |
|---|---|---|
| $f$ | Hz | Grid frequency |
| $I_1$ | A | Electric current phase L1 |
| $I_2$ | A | Electric current phase L2 |
| $I_3$ | A | Electric current phase L3 |
| $U_1$ | V | Voltage of phase L1 |
| $U_2$ | V | Voltage of phase L2 |
| $U_3$ | V | Voltage of phase L3 |
| $P_1$ | W | Electric power phase L1 |
| $P_2$ | W | Electric power phase L2 |
| $P_3$ | W | Electric power phase L3 |
| *$W_1$ | kWh | Energy phase L1 |
| *$W_2$ | kWh | Energy phase L2 |
| *$W_2$ | kWh | Energy phase L3 |
| $PF_1$ | — | Power factor phase L1 |
| $PF_2$ | — | Power factor phase L2 |
| $PF_3$ | — | Power factor phase L3 |
| $P$ | W | Total electric power |
| $Q$ | var | Total reactive power |
| $PF$ | – | Total power factor |
| $W_{in}$ | kWh | Electric energy consumed |
| $W_{out}$ | kWh | Electric energy delivered |
| $WQ_{in}$ | kvarh | Reactive energy consumed |
| $WQ_{out}$ | kvarh | Reactive energy delivered |
| $W$ | kWh | Total active energy |
| $WQ$ | kWh | Total reactive energy |

**Table 5.** Measurements collected from electricity meters. Measurements marked with * are only available for Janitza meters.

| Measurement | Unit | Description |
|---|---|---|
| AH | g/m$^3$ | Absolute humidity |
| $Dc$ | ° from North | Current wind direction |
| $Dp$ | °C | Dew point |
| $H$ | kJ/kg | Specific enthalpy |
| $I_{gc}$ | W/m$^2$ | Global horizontal irradiance |
| $I_{gm}$ | W/m$^2$ | Mean global horizontal irradiance (10 min moving average) |
| $P_a$ | hPa | Ambient air pressure |
| $\varrho$ | g/cm$^3$ | Actual air density |
| $S_c$ | m/s | Current wind speed |
| $T_a$ | °C | Ambient air temperature |
| $U_a$ | % | Relative humidity of ambient air |

**Table 6.** Measurements collected from the weather station.

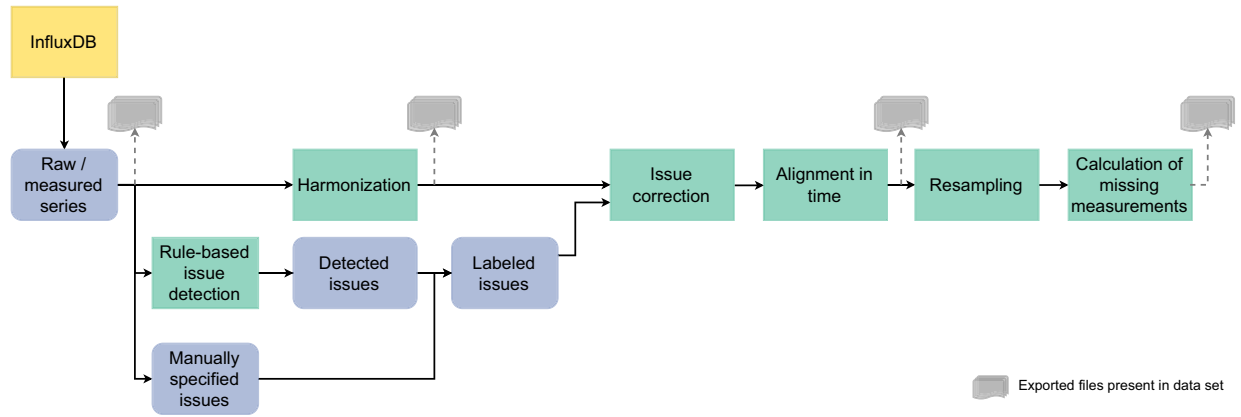

**Fig. 4** Data cleaning and post-processing pipeline for creating the presented dataset.

$$|\Delta W_{\text{meter}} - \Delta W_{\text{clone}}| < 0.9 \cdot |\Delta W_{\text{clone}}|, \quad (7)$$

the replacement $P(t)$ series as well as the integrated replacement energy series

$$W_{\text{clone}}(t) = W_{\text{meter}}(t_{\text{start}}) + \int_{t_{\text{start}}}^{t} P(\tau)\, d\tau \quad (8)$$

will be scaled and offset, such that

$$W_{\text{clone,adjusted}}(t_{\text{end}}) = f \cdot W_{\text{clone}}(t_{\text{end}}) + c = W_{\text{meter}}(t_{\text{end}}). \quad (9)$$

If instead $\Delta W_{\text{meter}} = 0$, the replacement power series will be copied without scaling. In all other cases, the replacement series will be copied, then scaled with a heuristic scaling factor

$$f = 1 - \left|\frac{\Delta W_{\text{meter}}}{\Delta W_{\text{clone}}}\right|, \quad (10)$$

which accounts for effects in the energy measurements caused by some meters/gateways after an outage. Note that in this case, $\Delta W_{\text{meter}}$ is usually significantly smaller than $\Delta W_{\text{clone}}$. The energy time series following a gap will be adjusted by offsetting such that it matches the value at the end of the gap after correction. The combined power-energy correction is applied to all power-energy measurement pairs, except for reactive power/energy of Janitza and Socomec meters, as reactive energy was erroneously recorded as integer instead of floating point values, hence joint correction was not reliable. Instead, the original clone gap filling mechanism described previously is applied.

Substitution for `V.K21`.  As described in the manual issue specification, the main cooling meter `V.K21` did not record reliable measurements for an extended amount of time due to mechanical failure. Its measurements however can be reconstructed from the downstream meters `H1.K11`, `H1.K12`, `H1.K14` and `H1.K16` (see Fig. 2). Temperature measurements are taken from `H1.K16`, and all flow related measurements are reconstructed by summing across all submeters, and scaling the resulting values to account for losses incurred between the physical locations of `V.K21` and the submeters.

Current transformer conversion factor corrections.  The PV meters and transformers are measured using clamped current transformers. After initial installation of the meters `H2.Z311`, `H4.ZE50` and `H4.ZE51`, the conversion factor of the corresponding current transformer was configured incorrectly, as mentioned above. As a correction, current, power and energy measurements are scaled accordingly to match the correct conversion factor.

*Alignment of measurements in time.*  After initial installation of the metering infrastructure in the facility, not all gateways, which assign timestamps to meter measurement data, had automatic time base synchronization using a time server due to technical restrictions. Therefore, a time drift in the measurement timestamps of the non-synchronized gateways is present, until synchronization was enabled for all gateways in March 2020. However, VisualGateway gateways had always been time synchronized, hence measurements collected through them can serve as a time basis for realigning drift-affected measurements. This is done based on measurements of the grid frequency, as its transient behavior is roughly equivalent across all meters in the facility, and can thus be used as a reference signal for synchronization. Since thermal meters and electrical meters of the same components are queried by the same gateway, both are affected by the same time drift, so the electricity meters' frequency measurements can be used to align thermal measurements. To align measurements in time, first, a reference frequency baseline is generated by averaging the frequency measurements across multiple VisualGateway-monitored meters. Then, for all non-synchronized gateways, the time deviation of a representative meter's frequency measurements from the reference baseline is derived by calculating the time difference between the daily extrema (min/max values) between measurement and reference. From these deviations, a time correction curve is determined for each gateway. This time correction curve is applied to all measurements of all meters for each gateway accordingly. Eventually, the underlying issue was resolved by enabling periodic time synchronization for all gateways.

*Data resampling.* Since most data is collected at a non-equidistant sample rate due to change-of-value measurements and time drift adjustment, the processed data is resampled into equidistantly sampled time series with 1 min resolution. This is generally done using linear interpolation. However, if the interval between a sample point and the nearest measurement value exceeds a threshold of 5 min, instead of linear interpolation, forward-fill of the last known measurement is applied. Note that this does not apply to weather station measurements, as they are affected by long gaps of missing measurement data which were not filled (as described previously). Instead of applying forward-fill, the gaps are filled by `NaN`. For all measurements, additional time series with 15 min and 1 h resolution are derived from this equidistantly sampled time series by downsampling. For non-cumulative measurements, the mean value of the measurements between two sample periods is used. Cumulative measurements, i.e., energy and flow volume measurements, are downsampled by linear interpolation.

*Calculation of missing measurements.* Contrary to all other electricity meters in facility, the ABB-B24 meters do not collect a direct $W$ or $WQ$ measurement, only $W_{in}/WQ_{in}$ and $W_{out}/WQ_{out}$ measurements are collected. Therefore, after all processing steps have been applied, these missing measurements are calculated as $W = W_{in} - W_{out}$ and $WQ = WQ_{in} - WQ_{out}$ respectively. A special case is the PV meter `V.Z84`, where $W$ is instead set to $W = W_{out}$, due to unreliable $W_{in}$ measurements. As the meter only measures PV power production and does not measure any consumers, the resulting error is negligible. Data from ABB-B24 meters has been collected as change-of-value, therefore $W_{in}/WQ_{in}$ and $W_{out}/WQ_{out}$ have differing start and end times, depending on the first and last value recorded. Hence, before calculating $W$ via the balance equation, for both time series, the first recorded value has been back-filled to the start of the dataset time frame, and the last recorded value forward-filled until the end of the dataset time frame.

*Export of processed data.* Eventually, the raw, harmonized, as well as processed and resampled time series are stored separately as compressed CSV files. Furthermore, all labeled issues are exported. The resulting data records are described in the following section.

## Data Records

**Full dataset.** The presented dataset contains raw and processed time series for each measurement, as listed in Tables 4, 5 and 6 respectively, for all meters listed in Tables 2 and 3. The dataset contains measurements from January 1, 2018 0:00 GMT+1 until January 1, 2024 0:00 GMT+1. Note that the facility is located in Offenbach, Germany, hence the local timezone is Europe/Berlin, which corresponds to GMT+2 during the European daylight savings period, and GMT+1 in winter. The dataset has a total size of 103 GB and is available on Dryad[27]: https://doi.org/10.5061/dryad.73n5tb363.

The main part of the dataset is the directory containing the time series measurement data of all listed meters. Additionally, the dataset contains a reduced dataset in a separate directory. This reduced dataset set contains a subset of the full set of meters and is described in the following subsection. The directory containing time series data consists of one directory for each listed meter, named by its URN, which in turn contains multiple time series for each measurement at different processing steps, as gzip-compressed CSV files, namely

- `URN_MEASUREMENT_raw.csv.gz`: raw, unprocessed time series (not present for measurements which are to be renamed during harmonization or calculated in the final processing step),
- `URN_MEASUREMENT_harmonized.csv.gz`: time series with applied harmonization step (not present for measurements calculated in the final processing step),
- `URN_MEASUREMENT_corrected.csv.gz`: time series with applied harmonization, issue correction and time alignment (not present for measurements calculated in the final processing step),
- `URN_MEASUREMENT_corrected_resampled_{1min|15min|1h}.csv.gz`: fully processed time series, resampled to 1 min, 15 min and 1 h resolution, respectively.

Note that in the non-harmonized raw series, the units as listed in Tables 4, 5 and 6 may not apply to all meters. Each file contains two columns of data, as shown in Table 7. Furthermore, each measurement of each meter may have individual start and end times of its resampled measurement time series, due to the change-of-value data collection. The time series start with the first recorded sample, binned to the closest minute. The end time likewise corresponds to the last sample recorded before January 1, 2024 0:00 GMT+1, binned to the closest minute. All resampled time series however have coinciding sample periods, i.e., full minutes.

Our motivation is to cater to different communities potentially interested in the published dataset. On the one hand, raw data may be of interest to practitioners focused on research questions from the energy management domain, on the other hand, corrected data allows for easier application of machine learning algorithms, i.e., without the need to extensively clean the data. We would like to emphasize again that the data stems from a highly complex real-world environment that is prone to measurement and other errors in the recorded data. The corrections applied aim at correcting these issues to provide a consistent and realistic dataset.

| Column | Unit | Description |
|---|---|---|
| datetime_utc | ISO 8601 string with time zone information | Measurement timestamp in UTC |
| *URN.measurement* | see Tables 4, 5 and 6 | Value of the given *measurement* of the meter *URN* |

**Table 7.** Data columns present in the files of the full dataset.

**Reduced dataset.** In addition to the full dataset, we provide a reduced dataset with a less complex representation of the building energy consumption, production, and weather measurements. This reduced data has a size of 320 MB in compressed form. The dataset is separated into 4 categories, as shown in Table 8. For electricity, heating and cooling, both power ($P$) and energy ($W$) aggregations are provided. The aggregations are generated by summing the fully processed $P$ and $W$ measurements, respectively, of the listed URNs. The units of all aggregations are identical to the original units as provided in Tables 5 and 6. All aggregations are provided from January 1, 2018 0:00 GMT+1 until January 1, 2024 0:00 GMT+1. The aggregated dataset is provided with 1 min, 15 min, and 1 h sample resolution. For each resolution, one sub-directory is provided, containing 7 gzip-compressed CSV files, namely

- `electricity_{P|W}.csv.gz`,
- `heating_{P|W}.csv.gz`,
- `cooling_{P|W}.csv.gz`,
- `weather.csv.gz`.

Each file contains a date-time column, analogous to Table 7, and one column per entry listed in Table 8. For each category, i.e., heating, cooling, electricity, and weather, additional plots are included in a separate directory in the dataset, illustrating the main measurements over the full time span.

**Labeled issues.** Furthermore, the data contains all manually and automatically detected issues for each of the meters present in the dataset. Manually specified and automatically detected issues are stored in two respective directories, `manual_issues` and `automatic_issues`. In each directory, all issues belonging to one respective meter are contained in a YAML file named `URN_issues_manual/automatic.yaml`. An annotated exampled of a labeled issue is available in the GitHub repository referenced in the code availability section.

## Technical Validation

**Validation of data consistency.** Before applying the data processing pipeline outlined in the methods section, all measurements of every meter listed in Tables 2 and 3 were inspected visually over the full 6-year time span. In addition to this, the unprocessed dataset had been used for monitoring purposes at the facility. After applying the data processing pipeline, the main measurements of the main meters (transformers, PV, heating/cooling) were inspected visually. The dataset was furthermore checked for consistency in an automatic fashion by ensuring all listed measurements were available, all gaps had been corrected, and the post-processed time series matched in length.

Table 9 summarizes the detected and manually specified issues in the dataset. The majority of issues are data gaps. While gaps are frequent, their average duration is relatively short, around 1 hour, though this value is skewed by longer gaps caused by gateway failures (see usage notes section) and extended meter outages, such as of meter `V.K21`. Overall, approximately 3.1% of the total time is affected by gaps, which is within acceptable limits given the dataset's size and sensor count. There is a significant variability across meters, with some experiencing frequent or prolonged gaps and others exhibiting minimal or no issues. Leaps in measurements are very rare, as expected, and occurrences of zero measurements, while more common, remain low and affect only single time steps, having minimal impact. Overall, the dataset contains an acceptably low level of issues, and we conclude that correcting these is unlikely to compromise its reliability.

**Building energy consumption characteristics.** To further validate the dataset, we analyzed the electrical energy production and consumption characteristics. Figure 5 presents a Sankey diagram illustrating the electrical energy flows in the facility over the full 6-year measurement period. The majority of energy production (62%) is sourced from the power grid. Approximately 16% are attributed to the PV system and 22% to the CHP. Additionally, about 2% of the energy is fed back into the grid during periods of excess PV power availability. The largest consumers of energy are the emission lab and workshops, together making up about 60% of the total electrical energy consumption. About 13% are attributed to cooling and ventilation of the facility. The remaining 27% of energy use are attributed to office spaces and various equipment.

Figure 6 depicts the energy consumption and production for each of the six years in the dataset. Overall energy consumption has decreased by approximately 13% since 2018, likely influenced by the COVID-19 pandemic and the increased prevalence of remote work in 2020 and 2021 and thereafter. Energy consumption from the grid has reduced significantly, by 52%, primarily due to the installation of the PV system. Electricity production by the CHP has fluctuated over the years, driven by variations in winter severity and associated total heating demand as the CHP operates only when there is sufficient heating demand. Given a power-to-heat ratio of 0.677, the CHP served 37% (2018) to 52% (2023) of the total heat demand. The significant increase in share of the total heat demand in 2023 is due to modernizations of the heating and ventilation system of the facility. As part of these modernizations, the integration and control of the CHP was adapted, allowing for a more efficient use of the CHP's capacities, further reducing the overall grid consumption. Notably, 2022 recorded the highest energy consumption for cooling, likely due high ambient temperatures for extended periods of time in summer, as illustrated in Fig. 7.

| Measurement | Description | URNs |
|---|---|---|
| **Electricity** | | |
| `total` | Electricity drawn from the main grid | V.Z81, V.Z82, H2.Z35, H2.Z36, H2.Z351, H2.Z361 |
| `PV` | PV production | V.Z84, H1.Z310, H1.Z311, H1.Z312 |
| `CHP` | CHP electricity production | H1.Z20 |
| **Heating** | | |
| `total` | Total heat production | H1.W11 |
| `CHP_heat` | CHP heat production | H1.W12 |
| `CHP_elec` | CHP electricity production | H1.Z20 |
| **Cooling** | | |
| `total` | Total cooling production of the cooling machines | V.K21 |
| `cool_elec` | Electricity consumption of the cooling machines | H1.Z16, H1.Z11, H1.Z12, H1.Z24, H1.Z25 |
| **Weather** | | |
| `Igm` | Mean global horizontal irradiance | WeatherStation.Weather |
| `Ta` | Ambient temperature | WeatherStation.Weather |

**Table 8.** Aggregated data provided in the reduced dataset. The data is aggregated by summing the corresponding measurements over all listed URNs.

| Category | Total Number of Issues | Average Duration (in s) | Total time ratio (in %) |
|---|---|---|---|
| Gap | 142,207 | 3283.47 | 3.12 |
| Lasting leap | 1 | — | — |
| Single leap | 3 | — | — |
| Zero measurement | 6,752 | 60.00 | 0.00 |

**Table 9.** Statistics on the types of issues automatically detected and corrected in the dataset.

Figure 7 further illustrates the energy consumption and production within the industrial facility, as well as corresponding weather measurements over the full dataset period. The upper panel shows the electrical load consumption alongside power production of the PV system and CHP. The electrical load consumption is calculated from the sum over all transformers, subtracted by the power production of the PV system and the CHP. The building's electrical load follows a similar yearly pattern across most of the observed measurement period. In 2023, while the overall electrical consumption reduced slightly in comparison to the previous years, a change in consumption patterns can be observed, which is caused by increased use of certain test equipment installed at the facility. A distinct reduction in consumption is visible during the company's Christmas break when operations are paused, leading to a static and low electricity demand. The COVID lockdown, as listed in Table 10, and the following change in office occupancy caused no notable differences in electrical or thermal demands, apart from a lower total electrical energy consumption. Further, the PV system's expansion can be observed, as outlined in Table 10. The small PV system in the parking lot became operational in 2019, while the larger PV installation was commissioned in mid-2020. The CHP system primarily operates during the colder months, aligning with the facility's heating demand shown in the middle panel. In the summer of 2023, the effect of the aforementioned change of the CHP operation can be observed, as the CHP continues to operate throughout the warmer months of the year.

The middle panel of Fig. 7 illustrates the total heating and cooling load. The heating load represents the thermal energy supplied by both the CHP and the gas boiler, i.e., meter `H1.W12`. The cooling load accounts for the total thermal cooling power produced by the three central cooling machines and measured by the central cooling meter `V.K21`. While the physically effective thermal power of the cooling machines is negative and thus plotted as such, the measurements of thermal power found in the dataset follow the overall sign convention and have positive sign. Both heating and cooling loads exhibit seasonal patterns strongly influenced by ambient weather conditions. A notable peak in heating demand is observed during the exceptionally cold winter of 2021. Additionally, an anomaly in the cooling load is observed on September 20, 2023, between 18:00 and 06:00. During this anomaly, all main-level cooling meters, i.e., `V.K21`, `H1.K11`, `H1.K12`, `H1.K14` and `H1.K16` exhibit the same behavior: Despite the cooling machines not running, an anomalously high thermal power $P$ is measured, while all other measurements of the meters remain consistent, including the cumulative energy measurement $W$. While the power readings are anomalously high, they have transient behavior consistent with all other measurements of the meters, indicating that likely a temporary measurement scaling issue occurred in the meters. Since the readings are not exceeding physically possible readings, the automatic issue detection does not flag the period as an issue. Since the failure occurred almost simultaneously across multiple meters and all non-power measurements are consistent, we chose to not manually remove or manipulate the respective measurement period. The two other, lower peaks in cooling load in 2023 can be attributed to a similar anomaly occurring in the meters.

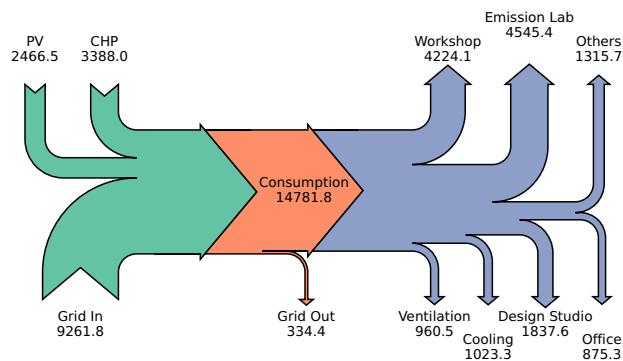

**Fig. 5** Sankey diagram showing the electrical energy flows of the facility over the measurement period of 6 years in MWh.

The lower panel presents the ambient air temperature and mean global horizontal irradiance. These measurements provide insight into the external environmental factors influencing energy demand and generation. Data gaps due to hardware failures of the weather station are visible in 2018; however, for the remainder of the dataset, only minor interruptions are present. The winter of 2020-2021 is identified as the coldest period in the dataset, corresponding with the high heating energy production observed in Fig. 6 for 2021 and the high heating load illustrated in the middle panel.

Figure 8 shows the facility's energy consumption and production for a representative week, specifically the first week of March 2021. This period, situated in early spring, is characterized by mild to low ambient temperatures and mostly clear skies. In contrast to the previous figure, the top panel shows the power drawn from and fed into the grid by the facility instead of the facility's electrical load, together with the PV system's and CHP's power production. During the weekend, the facility's overall electrical load significantly decreases. As a result, surplus PV power is available and is fed back into the grid. During the workweek, brief periods of excess PV power availability occur around noon, leading to grid export. This coincides with CHP operation, which becomes active after noon due to high heating demand during working hours. The CHP system operates primarily during the day, while its modulation strategy, which requires a minimum modulation threshold of 50% power, leads to on/off cycling during nighttime. The bottom panel shows the heating and cooling loads, analogous to Fig. 7, together with the ambient temperature. On the weekends, heating is shifted to morning and evening hours, causing the on/off cycling of the CHP to shift to the afternoon instead. A consistent, albeit small, cooling load is present throughout the week, primarily attributed to the emission lab cooling system (`H1.K11`) and the server cooling system (`H1.K16`). These subsystems require continuous cooling regardless of external conditions. Notably, the emission lab remains active over the weekend, necessitating light cooling even during non-working hours.

## Usage Notes

**General notes.** The aim in creating this dataset was to provide the research community with a consistent, curated, high-quality dataset. Since the data describes a complex, real-life system, it contains a set of issues which needed to be corrected. The issues most difficult to correct for, are gaps. In the methods section, we described the process we applied for filling these gaps using a heuristic copy-and-scaling mechanism. This process yields realistic and smooth data. Since this correction mechanism introduces artificial data, both raw data and the code for applying issue corrections have been made available. Furthermore, in principle, the meter hierarchy as illustrated in Fig. 1 could be used to correct issues across meters. However, we did only make use of this possibility for the correction of the central cooling meter `V.K21` as described earlier.

**Meter replacement of office transformer meters.** The meters of the office transformers `H2.Z35` and `H2.Z36` were decommissioned and replaced by new Janitza meters with distinct URNs `H2.Z351` and `H2.Z361`, respectively. The meters `H2.Z35/36` were decommissioned on September 9, 2020 at 12:00 UTC. The new meters `H2.Z351/Z361` were commissioned on September 15, 2020 at 10:00 UTC. Between these two dates, no data for the office transformers is available.

**Weather station.** Due to repeated hardware failure throughout the dataset's measurement period, no weather data is available for extended periods of time (up to multiple days). These periods are included in the automatically detected issues. While we chose not to correct these issues automatically, data from a nearby weather station is available from the German Meteorological Service (https://www.dwd.de/EN/ourservices/cdc/cdc_ueberblick-klimadaten_en.html, station ID 07341).

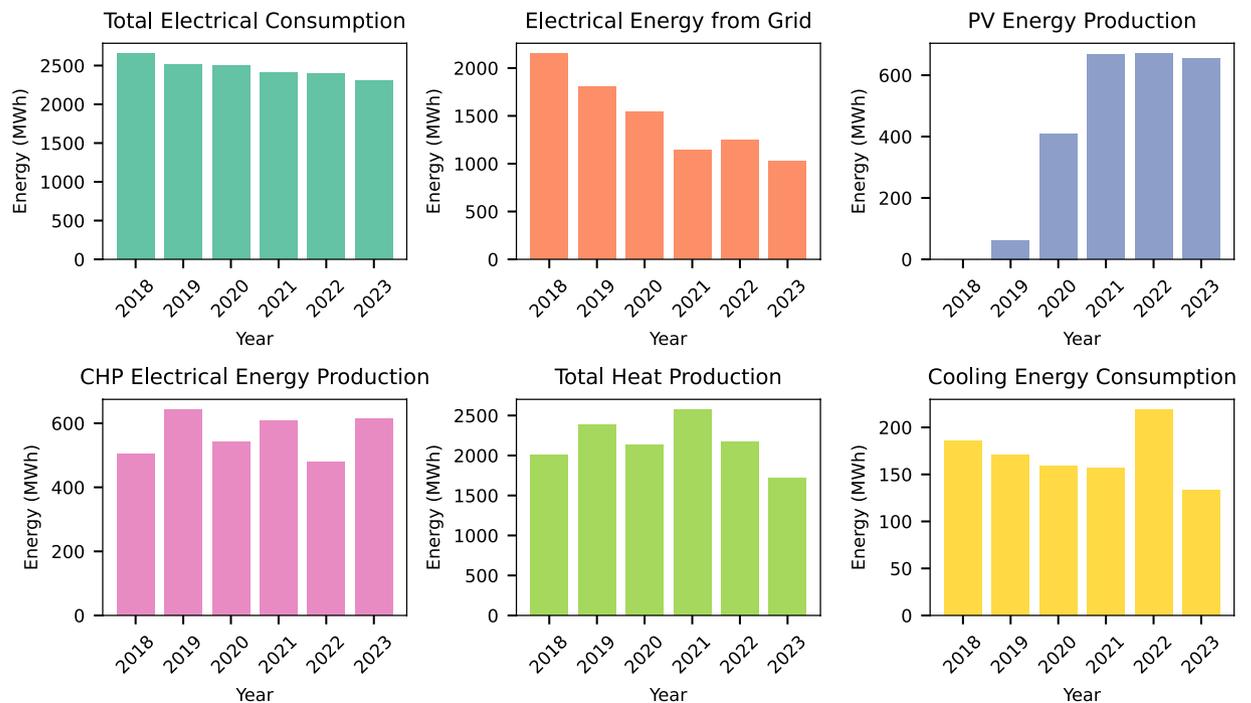

**Fig. 6** Total electrical and thermal energy consumption and production over the course of the 6 years present in the dataset.

**PV system.** The PV system configuration has changed during the recording of the dataset. The first smaller system was installed in June 2019 and in June 2020 an upgrade to its final capacity, which is described in the facility description, was made. This is noted in Table 10 and illustrated in Fig. 7.

**CHP.** Before February 13, 2019, the CHP could only be operated at 100% or 0% capacity. After this date, the CHP's control logic was updated to allow for a minimal capacity of 50% and linear modulation between 50% and 100% capacity. Even after this update, the CHP's capacity was not fully utilized, due to how it was integrated in the facility's heating system. This issue was addressed as part of a modernization of the overall heating and ventilation system in June 2023, enabling the CHP to fulfill a significant share of the overall heating demand without changes to the modulation scheme.

**Known gateway failures.** On four occasions, gateways collecting measurement data from meters failed for extended periods, resulting in missing measurement data across all meters queried by the respective gateways. In our setup, the gateways correspond to the subdivisions illustrated in Fig. 1, with the exception that Janitza meters are queried by a separate gateway, which was not affected by any outages. The following gateway failures are present in the data:

- First workshop gateway failure from 2020-02-13 until 2020-03-06
- Emission lab gateway failure from 2020-08-20 until 2020-09-17
- Distribution gateway failure (meters `H2.T.Z30 - H2.T.Z32, H2.K21`) from 2021-11-15 until 2021-12-10
- Second workshop gateway failure from 2022-05-06 until 2022-07-14

The gateway failures were caused by hardware defects of Tixi Data Gateways, which subsequently had to be replaced with new units. The failures were confined to single gateways, meaning that all metering on levels above the affected subdivisions were not affected. The listed gateway failures were detected using the automatic issue detection and corrected as described in the methods section.

**P and W measurements.** The power time series $P$ reflect measurements of the instantaneous power at sample time of each specific meter, while the corresponding energy time series $W$ record the accumulated energy up to sample time. In theory, the power signal $P$ is given by the derivative of the energy signal $W$ with respect to time. However, for discretely sampled time series this is not true and the numerical derivative of $W$ with respect to time does not reproduce $P$. Instead, it only gives the average power over the sample interval. Similarly, integrating the power series $P$ over time will in general not exactly reproduce the energy series $W$. These discrepancies will be most prominent for fast switching devices. However, if the sample frequency of measurements in $P$ is sufficiently high, the corresponding errors will be small. A further source for discrepancies between both values are the issue correction mechanisms, especially the gap correction and resampling of the non-equidistant change-of-value time series into equidistantly sampled time series. In the provided dataset, the conditions for obtaining small errors are mostly met. Overall, we have found that $\int_{t_0}^{t_1} P(\tau)\, d\tau \approx W(t_1) - W(t_0)$ with an average absolute relative error over all time series of the dataset of approximately 0.28%. Generally, the error of the numerical derivative of $W$ is much larger due to the fact that finite differences are much more sensitive to small changes in the data. We explicitly show two examples comparing the measured power series $P$ with numerical estimates of $\frac{dW}{dt}$ with finite differences, i.e., $\frac{\Delta W}{\Delta t}(t_i) = \frac{W(t_{i+1}) - W(t_i)}{\Delta t}$, in Fig. 9. While the left example for the meter H1.Z15 shows good agreement, the second example of H1.Z20 shows larger discrepancies. Caution is therefore advised when choosing and comparing power and energy domains.

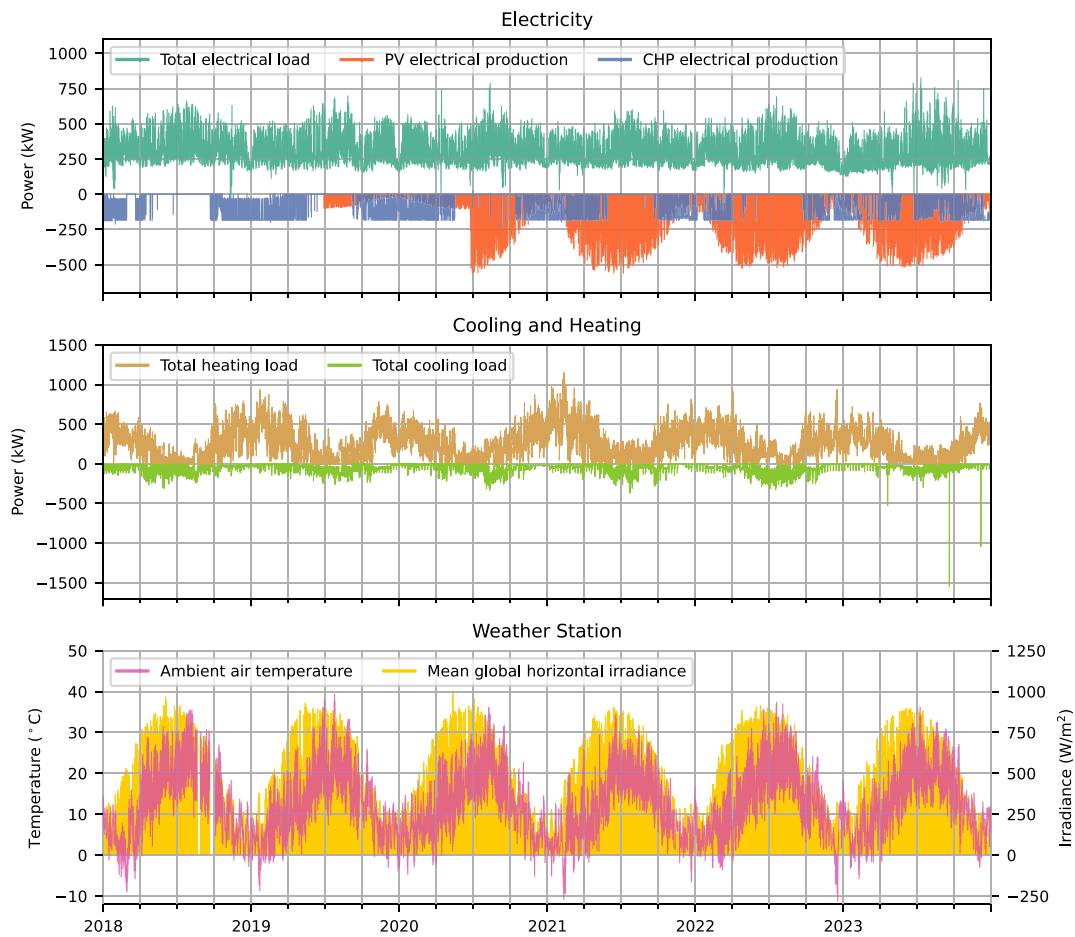

**Fig. 7** Energy consumption, production, and weather conditions over the full dataset period. The upper panel displays electrical load consumption alongside power production from the PV system and CHP. The middle panel shows heating and cooling loads, including an observed anomaly in cooling meter readings in 2023. The lower panel presents ambient temperature and global horizontal irradiance as measured by the weather station.

| Event | Time |
|---|---|
| Update of CHP control mode | February 19, 2019 |
| Commissioning of parking lot PV plant (groups 1, 2) | June 29, 2019 |
| Commissioning of rooftop PV plant installation (groups 4–6) | June 25, 2020 |
| Decommissioning of old office transformer meters | September 9, 2020 |
| Commissioning of new office transformer meters | September 15, 2020 |
| COVID lockdown | March 16, 2020 - January 17, 2021 |
| Modernization of the heating system and CHP integration | June 2023 |

**Table 10.** Important events affecting the facility during the measurement period.

**Relation of measurement values.** For all but the ABB-B24 meters, $W/WQ$, $W_{in}/WQ_{in}$ and $W_{out}/WQ_{out}$ are separately measured quantities, that further undergo decoupled issue correction. Therefore, in both the raw and processed data, the relations $W = W_{in} - W_{out}$ and $WQ = WQ_{in} - WQ_{out}$, respectively, only hold in approximation. As described in the issues correction, however, for ABB-B24 meters it does hold, since $W/WQ$ is calculated during post-processing. Similarly, total and phase-wise power measurements $P$, $P_1$, $P_2$ and $P_3$ are measured and corrected separately, so $P \approx P_1 + P_2 + P_3$ for non-Janitza meters. For Janitza meters, the total power $P$ and energy $W$ is calculated internally by the meter from the phase-wise energy, thus $P = \sum_{i=1}^{3} P_i$ and $W = \sum_{i=1}^{3} W_i$ in the raw time series. However, all measurements are corrected separately, so this relation only holds in approximation after correction.

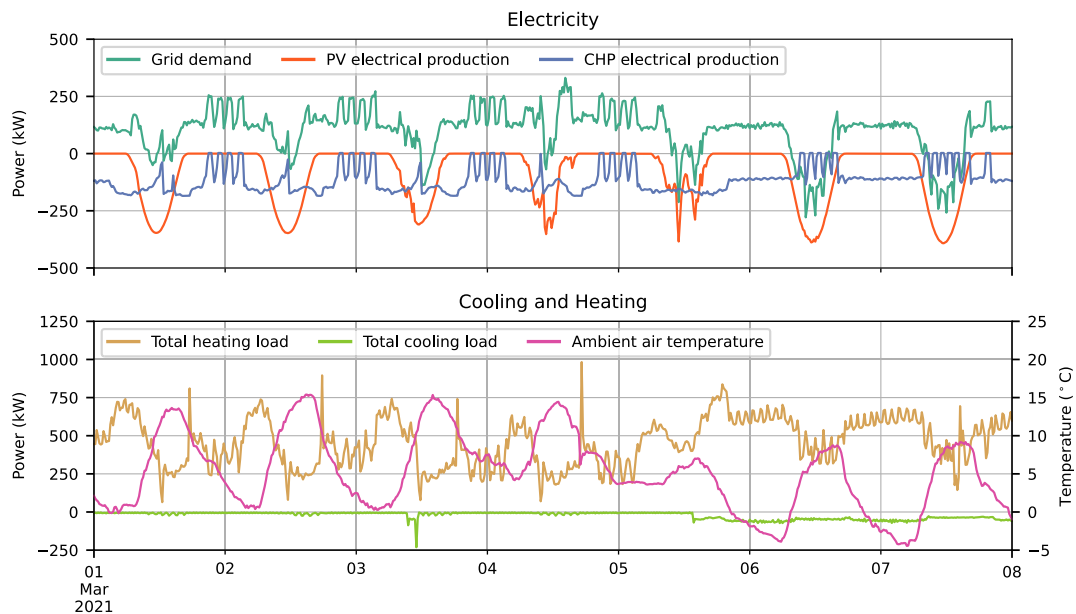

**Fig. 8** Representative time series of the facility's energy consumption and production for the first week of March 2021. The top panel illustrates grid demand, i.e., the power drawn from and fed into the grid, along with PV and CHP power production. The bottom panel shows heating and cooling loads along with ambient temperature.

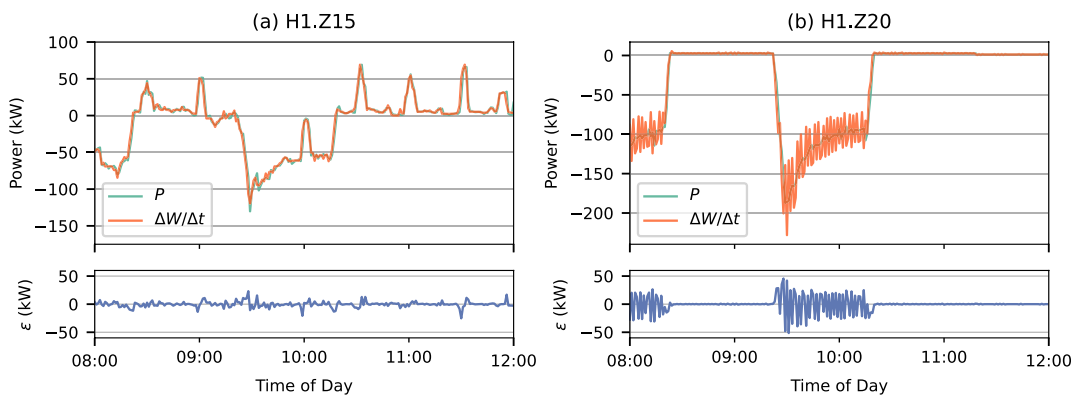

**Fig. 9** Comparison of instantaneous power measurements $P$ and numerical derivatives of $W$ for two different meters during an exemplary measurement period. The bottom plot in each panel shows the deviation $\epsilon$ between $P$ and the numerical derivative of $W$, i.e., $\epsilon(t) = P(t) - \frac{\Delta W}{\Delta t}(t)$.

**Issues in recording of *WQ* measurements.** Due to faulty gateway configurations, the measurements of the total reactive energy *WQ* of Socomec and Janitza meters were recorded with integer precision instead of floating point precision.

**Defective meters `H4.Z50/51`.** Since the end of June 2023, the meters `H4.Z50` and `H4.Z51` became unavailable due to hardware failure. This outage period has not been filled with reconstructed data. However, replacement data is available from meters `H4.ZE50` and `H4.ZE51`, as indicated in Table 3.

**Redundant metering.** Some components in the building have been metered redundantly since the installation of secondary meters in 2023 as indicated with `ZE` in Tables 2 and 3 and illustrated in Fig. 1. While this redundant information could be leveraged for issue correction, this has not been done in the correction pipeline, instead both Z and ZE meters have been processed separately. However, as illustrated in the previous note, the availability of both sets of measurement data can still be helpful for further analyses.

**Unaddressed issues.** In some rare cases, not all measurements from a given meter are present in the dataset due to the batch-wise querying process, where different measurements from the same meter are queried in separate batches. As a result, if an outage occurs between batches, individual measurements may be lost without affecting others from the same meter. However, the gap detection mechanism operates at the sensor level and assumes that outages impact all queried measurements simultaneously, focusing primarily on the main measurements (e.g., energy or power). Consequently, if a non-main measurement is missing while the main measurements used for issue detection are successfully queried, the data gap is not detected or corrected.

## Code availability

The Python code for reproducing the technical validation results, i.e., figures, tables and statistics, as well as the code for downsampling the corrected measurement time series, and for generating the reduced aggregated dataset are available on GitHub at https://github.com/HRI-EU/MonitoringDatasetAnalysis.

### Acknowledgements
The authors thank Martin Stadie and Sven Rebhan for their support, as well as Monika Wicke and Markus Mauksch for their diligent work in supporting the creation of the dataset. S.S. acknowledges funding by the European Union. Views and opinions expressed are however those of the author(s) only and do not necessarily reflect those of the European Union or European Commission. Neither the European Union nor the granting authority can be held responsible for them. Grant Agreement 101080086 – NeQST.

### Author contributions
CRediT: Conceptualization: J.E., A.C., P.W., T.R.; Data curation: J.E., A.C., S.S., R.U.; Formal Analysis: J.E., A.C., T.S., S.S., R.U.; Investigation: J.E., A.C., F.L., T.S., S.S., L.F., S.L., R.U.; Methodology: J.E., A.C., P.W., S.S., R.U.; Project administration: PW, T.R.; Resources: D.L., F.J., R.U.; Software: J.E., A.C., T.S., S.S., R.U.; Supervision: P.W., T.R.; Validation: J.E., A.C., S.S.; Visualization: J.E., A.C., F.L., T.S., L.F.; Writing - original draft: J.E., A.C., S.S.; Writing - review & editing: J.E., A.C., P.W., F.L., T.S., S.S., L.F., S.L., D.L., F.J., R.U., T.R.

### Competing interests
JE, AC, PW, FL, TS, SS, LF, SL, DL and TR are employees of Honda Research Institute Europe. FJ is an employee of Honda R&D Europe (Germany). RU is an employee of EA Systems Dresden and completed consultancy work paid by Honda Research Institute Europe as part of the data acquisition and processing for this study. Funding parties had no role in the collection, analysis, and interpretation of data, and had no role in the decision to publish, or the preparation of the manuscript.

### Additional information
**Correspondence** and requests for materials should be addressed to J.E.

**Reprints and permissions information** is available at www.nature.com/reprints.

**Publisher's note** Springer Nature remains neutral with regard to jurisdictional claims in published maps and institutional affiliations.